\documentclass[sigchi]{acmart}
\usepackage{color,soul}
\usepackage{makecell}
\def\BibTeX{{\rm B\kern-.05em{\sc i\kern-.025em b}\kern-.08emT\kern-.1667em\lower.7ex\hbox{E}\kern-.125emX}}

\begin{document}

%
% The "title" command has an optional parameter, allowing the author to define a "short title" to be used in page headers.
\title{The EMPATHIC Project: Mid-term Achievements}

%
% The "author" command and its associated commands are used to define the authors and their affiliations.
% Of note is the shared affiliation of the first two authors, and the "authornote" and "authornotemark" commands
% used to denote shared contribution to the research.
%\author{Ben Trovato}
%\authornote{Both authors contributed equally to this research.}
%\email{trovato@corporation.com}
%\orcid{1234-5678-9012}
%\author{G.K.M. Tobin}
%\authornotemark[1]
%\email{webmaster@marysville-ohio.com}
%\affiliation{%
%  \institution{Institute for Clarity in Documentation}
%  \streetaddress{P.O. Box 1212}
%  \city{Dublin}
%  \state{Ohio}
%  \postcode{43017-6221}
%}

\author{M. I. Torres, J. M. Olaso}
\author{C. Montenegro, R. Santana}
\author{A. V\'{a}zquez, R. Justo}
\author{J. A. Lozano}
    \affiliation{%
    \institution{Universidad del Pa\'{i}s Vasco UPV/EHU}
    \city{Bilbao}
    \country{Spain}
}
\email{manes.torres@ehu.eus}

\author{S. Schl\"{o}gl}
\affiliation{%
    \institution{MCI Management Center Innsbruck}
    \city{Innsbruck}
    \country{Austria}
}
\email{stephan.schloegl@mci.edu}

\author{G. Chollet, N. Dugan, M. Irvine}
\author{N. Glackin, C. Pickard}
\affiliation{%
    \institution{Intelligent Voice Ltd}
    \city{London}
    \country{UK}
}
\email{gerard.chollet@intelligentvoice.com}
 
\author{A. Esposito, G. Cordasco}
\author{A. Troncone}
\affiliation{%
    \institution{Universit\`{a} degli Studi della Campania, Luigi Vinvitelli}
    \city{Caserta}
    \country{Italy}
}
\email{iiass.annaesp@tin.it}

\author{D. Petrovska-Delacretaz}
\author{A. Mtibaa, M. A. Hmani}
\affiliation{%
    \institution{Institut Mines-Telecom}
    \city{Evry}
    \country{France}
}
\email{dijana.petrovska@telecom-sudparis.eu}

\author{M. S. Korsnes}
\author{L. J. Martinussen}
\affiliation{%
    \institution{Department of Old Age Psychiatry, Oslo University Hospital}
    \city{Oslo}
    \country{Norway}
}
\email{m.s.korsnes@psykologi.uio.no>}

\author{S. Escalera}
\author{C. Palmero Cantari\~{n}o}
\affiliation{
    \institution{Universitat de Barcelona and Computer Vision Center}
    \city{Barcelona}
    \country{Spain}
}
\email{sergio@maia.ub.es}

\author{O. Deroo}
\author{O. Gordeeva}
\affiliation{
    \institution{Acapela Group}
    \city{Mons}
    \country{Belgium}
}
\email{olivier.deroo@acapela-group.com}

\author{J. Tenorio-Laranga}
\author{E. Gonzalez-Fraile}
\author{B. Fernandez-Ruanova}
\author{A. Gonzalez-Pinto}
\affiliation{%
    \institution{Osatek/ Osakidetza}
    \city{Bilbao}
    \country{Spain}
}
\email{jtenorio@osatek.eus}

% additional authors...

%
% By default, the full list of authors will be used in the page headers. Often, this list is too long, and will overlap
% other information printed in the page headers. This command allows the author to define a more concise list
% of authors' names for this purpose.
\renewcommand{\shortauthors}{M. I. Torres et al.}

%
% The abstract is a short summary of the work to be presented in the article.
\begin{abstract}
The goal of active aging is to promote changes in the elderly community so as to maintain an active, independent and socially-engaged lifestyle. Technological advancements currently provide the necessary tools to foster and monitor such processes. This paper reports on mid-term achievements of the European H2020 EMPATHIC project, which aims to research, innovate, explore and validate new interaction paradigms and platforms for future generations of personalized virtual coaches to assist the elderly and their carers to reach the active aging goal, in the vicinity of their home. The project focuses on evidence-based, user-validated research and integration of intelligent technology, and context sensing methods through automatic voice, eye and facial analysis, integrated with visual and spoken dialogue system capabilities. In this paper, we describe the current status of the system, with a special emphasis on its components and their integration, the creation of a Wizard of Oz platform, and findings gained from user interaction studies conducted throughout the first 18 months of the project.
%This paper reports on mid-term achievements of the European H2020 EMPATHIC project, which aims to research, innovate, explore and validate new interaction paradigms and platforms for future generations of Personalized Virtual Coaches to assist elderly people living independently in and around their home. Innovative multi-modal face analytics, adaptive spoken dialogue systems, and natural language interfaces are part of what the project investigates, aiming to help dependent aging persons and their carers. It uses remote, non-intrusive technologies to extract physiological markers of emotional states and adapt respective coach responses. In doing so, it develops causal models for emotionally believable coach-user interactions, which shall engage elders and thus keep off loneliness, sustain health, enhance quality of life, and simplify access to future telecare services. The project includes a demonstration and validation phase with clearly defined, realistic use cases. It focuses on evidence-based, user-validated research and integration of intelligent technology and context sensing methods through voice, eye and facial analysis, integrated with visual and spoken dialogue system capabilities. Through measurable end-user validations performed in Spain, Norway and France (and complementary user evaluations in Italy), the proposed methods and solutions will ensure usefulness, reliability, flexibility and robustness. This paper reports on mid-term achievements (after 18 months). 
\end{abstract}

%
% The code below is generated by the tool at http://dl.acm.org/ccs.cfm.
% Please copy and paste the code instead of the example below.
%



%
% Keywords. The author(s) should pick words that accurately describe the work being
% presented. Separate the keywords with commas.

%
% A "teaser" image appears between the author and affiliation information and the body 
% of the document, and typically spans the page. 
%\begin{teaserfigure}
%  \includegraphics[width=\textwidth]{Logos.png}
%  \caption{The EMPATHIC Virtual Coach Partners.}
%  \Description{Logos of the partners institutions.}
%  \label{fig:teaser}
%\end{teaserfigure}

%
% This command processes the author and affiliation and title information and builds
% the first part of the formatted document.
\maketitle

% - INTRODUCTION %%%%%%%%%%%%%%%%%%%%%%%%%%%%%%%%%%%%%%

\section{Introduction}

Despite advances in health care and technology, most of the elder care is still provided by informal caregivers, i.e. friends and family members. According to predictions, however, this type of care will decrease in the future, for which studies encourage society to concentrate on improving the elderly's lifestyle, helping them to remain independent for a longer period of time~\cite{Willcox2014HealthyAD}. In particular, the focus should be on external and internal difficulties of the elderly, offering arrangements and facilities to support active aging. Indeed, socio-behavioral and environmental conditions are a crucial factor affecting longevity~\cite{kirkwood2005}, which to some extent explains variations found in the aging process, ranging from active and positive to feeble and dependent. We believe that four principles promote active aging, namely dignity, autonomy, participation, and joint responsibility. Information and Communication Technologies (ICT) are expected to make such principles possible, allowing the elderly to stay active members of the societal community while helping them remain independent and self-sufficient~\cite{brinkschulte2018empathic}. 

Consequently, the EMPATHIC (\textit{Empathic, Expressive, Advanced Virtual Coach to Improve Independent Healthy-Life-Years of the Elderly}) project\footnote{http://www.empathic-project.eu/} aims to contribute to technological progress in this area by researching, innovating and validating new interaction paradigms and platforms for future generations of personalized Virtual Coaches (VC) to promote active aging. It is centred around the development of the EMPATHIC-VC, a non-obtrusive, emotionally-expressive virtual coach whose aim is to engage senior users in enjoying a healthier lifestyle concerning diet, physical activity, and social interactions. This way, they actively minimize their risk of potentially chronic diseases, which contributes to their ability to maintain a pleasant and autonomous life, while in turn it helps their carers. The main goal of the VC is to create a link between one's body and emotional well-being. To do so, it will perceive and identify users' social and emotional states by means of multi-modal face, eye gaze and speech analytics modules. Furthermore, it will learn and understand users' requirements and expectations, and adaptively respond to their needs through novel spoken dialogue systems and intelligent computational models. Such a combination of modules will allow for user-coach real-time interaction, thus promoting empathy in the user. 

In this paper, we describe our mid-term achievements, explaining where we currently stand with these goals, 18 months into the project. Section~\ref{sec:system} describes the current status of the system components, with particular emphasis on the most robust modules up to date. The integration of these modules is presented in Section~\ref{sec:technology}. Lessons learned from the preliminary human-coach interaction studies are explained in Section~\ref{sec:interaction}, while Section~\ref{sec:summary} concludes the paper.

% - SYSTEM COMPONENTS %%%%%%%%%%%%%%%%%%%%%%%%%%%%%%%%%%%%%%
\section{Status: System Components}
\label{sec:system}
The EMPATHIC VC is based on the following system components, each of which is researched, built and evaluated independently.  

\subsection{Automatic Speech Recognition}
The Automatic Speech Recognition (ASR) component turns speech from a continuous stream of audio into structured data, containing likely words and their alternatives, labelled with the confidence level for each, start time (as an offset from the stream start), and duration. So far, our main achievements with this component are:

\begin{itemize}

\item The development ASR.online, a new method of reading data into the ASR engine which uses the opensource GStreamer framework\footnote{https://github.com/GStreamer/gstreamer} to continuously stream the audio through the ASR executable. Compared with our previous approach of buffering audio and processing small chunks, ASR.online reduces the latency between the speech and the transcription. The average latency has been reduced to approximately 500 milliseconds, as a result of both overall performance improvements as shown in Table~\ref{table:asrperformance}, and the immediate transmission of transcription data. 

\item The training of acoustic models for French, Spanish and Norwegian, the languages which will be used in the EMPATHIC field trials. The acoustic models of the ASR component were trained using the Kaldi ASpIRE recipe\footnote{https://github.com/kaldi-asr/kaldi/tree/master/egs/aspire}.
The training data consists of 1067 hours of Spanish speech, 271 hours of French and 228 hours of Norwegian, augmented at the DNN-HMM training stage by the addition of noise from RWCP, AIR and Reverb2014 databases using the reverberation algorithm implemented in the Kaldi framework. The training process took approximately two weeks, and used two systems with 32 CPU cores and 64GB RAM each, with a total of 6 NVIDIA Pascal architecture GPUs. 3-gram language models were adapted using transcription data from the training set. These models contain a vocabulary of approximately 67,000 words in the Spanish model, 60,000 words in French and 48,000 in Norwegian. The Norwegian model uses Bokm{\aa}l, one of two official written standards of Norwegian. The models were tested using the NIST SCLITE utility\footnote{https://github.com/usnistgov/SCTK}, which scores the best-path output against the ground-truth for correct words, substitutions, deletions, insertions and an overall word error rate. The results are shown in Table~\ref{table:scliteresults}

\end{itemize}

\begin{table}[]
\caption {The relative performance of the original (ASR) and new (ASR.online) approaches.  All values are in seconds.}
\begin{tabular}{p{1.2cm} p{1.0cm} p{1.0cm} p{1.5cm} p{1.5cm}}
\hline
\makecell{Audio \\ Duration \\ (in sec.)} & \makecell{ASR \\ CPU \\ (in sec.)} & \makecell{ASR \\ GPU \\ (in sec.)} & \makecell{ASR.online \\ CPU \\ (in sec.)} & \makecell{ASR.online \\ GPU \\ (in sec.)} \\
\hline
\multicolumn{1}{c}{2.0} & \multicolumn{1}{c}{6.0} &\multicolumn{1}{c}{6.0}  & \multicolumn{1}{c}{2.5} & \multicolumn{1}{c}{1.5} \\
\multicolumn{1}{c}{5.0} & \multicolumn{1}{c}{7.0} & \multicolumn{1}{c}{6.0} & \multicolumn{1}{c}{3.0} & \multicolumn{1}{c}{1.5} \\
\multicolumn{1}{c}{10.0} & \multicolumn{1}{c}{7.0} & \multicolumn{1}{c}{6.0} & \multicolumn{1}{c}{4.6} & \multicolumn{1}{c}{2.0} \\
\multicolumn{1}{c}{15.0} & \multicolumn{1}{c}{10.0} & \multicolumn{1}{c}{6.0} & \multicolumn{1}{c}{6.5} & \multicolumn{1}{c}{2.5} \\
\multicolumn{1}{c}{20.0} & \multicolumn{1}{c}{13.0} & \multicolumn{1}{c}{6.0} & \multicolumn{1}{c}{8.0} & \multicolumn{1}{c}{4.0} \\
\multicolumn{1}{c}{25.0} & \multicolumn{1}{c}{15.0} & \multicolumn{1}{c}{6.0} & \multicolumn{1}{c}{9.5} & \multicolumn{1}{c}{4.5} \\
\multicolumn{1}{c}{40.0} & \multicolumn{1}{c}{22.0} & \multicolumn{1}{c}{8.0} & \multicolumn{1}{c}{17.0} & \multicolumn{1}{c}{5.5} \\
\multicolumn{1}{c}{70.0} & \multicolumn{1}{c}{35.0} & \multicolumn{1}{c}{9.0} & \multicolumn{1}{c}{26.0} & \multicolumn{1}{c}{5.5} \\
\multicolumn{1}{c}{100.0} & \multicolumn{1}{c}{47.0} & \multicolumn{1}{c}{10.0} & \multicolumn{1}{c}{38.5} & \multicolumn{1}{c}{7.0} \\
\multicolumn{1}{c}{130.0} & \multicolumn{1}{c}{63.0} & \multicolumn{1}{c}{12.0} & \multicolumn{1}{c}{45.0} & \multicolumn{1}{c}{7.5} \\
\multicolumn{1}{c}{600.0} & \multicolumn{1}{c}{279.0} & \multicolumn{1}{c}{25.0} & \multicolumn{1}{c}{210.0} & \multicolumn{1}{c}{16.5} \\
\multicolumn{1}{c}{1200.0} & \multicolumn{1}{c}{550.0} & \multicolumn{1}{c}{64.0} & \multicolumn{1}{c}{443.0} & \multicolumn{1}{c}{30.0} \\  
\hline
\end{tabular}
\label{table:asrperformance}
\end{table}

\begin{table}[]
\caption {SCLITE benchmarking for supported languages.}
\begin{tabular}{lccccc}
\hline
Language & {COR} & {SUB} & {DEL} & {INS} & {WER} \\
\hline
English & 87.7 & 9.2 & 3.1  & 4.0 & 16.3 \\
French & 73.9 & 22.7 & 3.4 & 10.1 & 36.3 \\
Spanish & 78.1 & 12.7 & 9.3 & 4.4 & 26.3 \\
Norwegian & 55.9 & 19.8 & 24.4 & 7.5 & 51.5 \\
\hline
\end{tabular}
\label{table:scliteresults}
\end{table}

\subsection{Natural Language Understanding}
The Natural Language Understanding (NLU) component translates the output of the ASR  into semantic units to be processed by the Dialog Management (DM) component. The main mid-term achievements in the conception of the NLU component concern the development of two multi-lingual methods for topic classification.

First we had to detect the user's end-of-turn pauses. Following previous approaches (e.g. \cite{roddy2018investigating,shannon2017improved}), we addressed this question as a classification problem. As features for classification, we use the temporal profile of speech, and the syntactic and semantic information encoded in the utterances. As classifiers, we used different variants of deep neural models. A distinguished feature of our approach is that we implemented an ASR simulator that allows us to evaluate the sensitivity of the End-of-Turn-Detection (EOTD) with respect to particular characteristics of the speaker (speech profiles), and to errors in the ASR output. This validation step is essential, since in the implementation of dialogue systems an early mistake in any of the system components can produce a negative cascading effect in the performance of the subsequent elements of the pipeline. %In the experiments conducted, the ASR simulator has shown to be a useful tool to characterize and measure the behaviour of the EOTD under different conditions.

Our NLU component is expected to treat user utterances in three + one different languages, i.e. Spanish, French and Norwegian + English. In the literature, such systems are usually referred to as multi-lingual models, and different strategies have been proposed to develop them. Learning becomes a challenging task for a multi-lingual system since languages are diverse in their grammar. Furthermore, the availability and quality of the corpora with which machine learning models are usually trained is not equal for the languages. Thus, we focused on topic classification. Our approach was to create a modular system where information available in one language is transferred or exploited while learning models for the other languages. We implemented two strategies: the first based on the use of the Wordnet semantic network and synsets \cite{Miller:1995}, the second based on parallel corpora.

Wordnet semantic synsets encapsulate information about different senses of commonly used words. This information is language independent, and can thus be used not only to obtain the equivalent word in another language, but also to obtain a set of sense-connected synsets, by computing the closure set that includes the hyperonymies. Thus, once we have a group of sense-connected synsets, we can generate the set of words that represent them for each language. %The first strategy we implemented uses this tool to generate a set of words for each of the topics we want to detect, used to label a corpus of human-to-human dialogues.

Our second strategy focuses on training one model for one particular language for which we have quality labeled data. By means of parallel corpora, we then extrapolate the obtained labels to a second language, so as to later train a specific model. Labeling is a tedious work, yet thanks to this strategy, we only have to label in one language while still obtaining a corpus for each of the languages.

\subsection{Dialogue Management} \label{sec:dm}
The Dialogue Management (DM) component maintains the state and manages the flow of the conversation or, in other words, determines the action a system has to perform at each step.

\begin{figure}[!t]
 \centering
 \includegraphics[width=\linewidth]{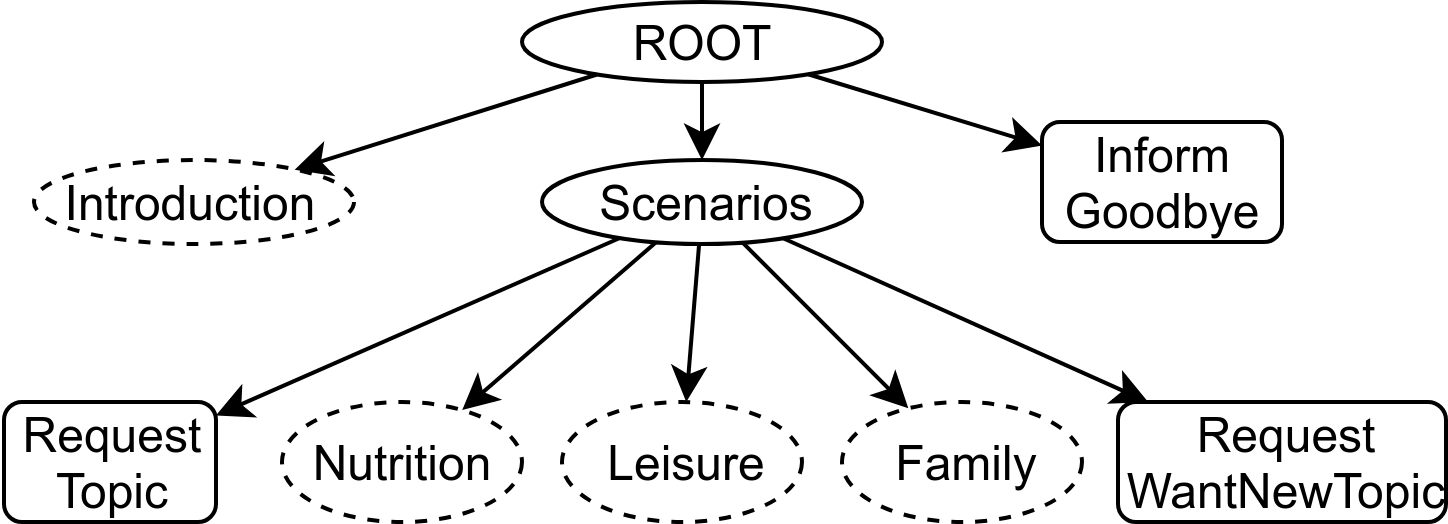}
 \caption{High level dialogue structure employed by the DM of the EMPATHIC VC.}
 \Description{High level dialogue structure empoyed by the DM of the EMPATHIC VC}
 \label{fig:dm}
\end{figure}

For the EMPATHIC VC we used a DM providing an advanced management structure based on distributed software agents. It enforces a clear separation between the domain-dependent and the domain-independent aspects of the dialogue control logic. The domain-specific aspects are defined by a dialogue task specification, and a domain-independent dialogue engine executes the given dialogue task to manage the dialogue. The dialogue task specification is defined by a tree of dialogue agents, where each agent is responsible for managing a sub-part of the dialogue. For instance, Figure~\ref{fig:dm} shows the high level dialogue task structure for the EMPATHIC VC where the ``Introduction'' agent is responsible for handling the introductory dialogue of the EMPATHIC VC, the ``Nutrition'' agent is responsible for handling the nutrition dialogues, etc.

The aim of the system is to have a virtual coach that helps users in certain aspects of their lives. To this end, we have emulated a GROW (Goal - Reality - Obstacle - Will)~\cite{Sayas2018} coaching model into the DM's dialogue strategy. The GROW model is a structured method based on problem solving, goal setting and goal-orientation. The Model is divided into four phases that propose four questions to guide the user towards obtaining and achieving a goal. These questions are asked in a pre-established order. In the first session, this order must be respected to facilitate the user to follow the thread and be able to explore its goal and the necessary steps to achieve it. In the following sessions, the order can be changed or specific phases  be chosen. Using the GROW dialogue model, the following dialogue topics have been implemented:

	\begin{itemize}
		\item \textbf{Introduction} dialogues to make the users feel comfortable with the system and to obtain some basic information. This dialogue is carried out the very first time a user dialogues with the system.
		\item \textbf{Sport and Leisure} \cite{Sayas_leisure2018} \cite{Sayas_physical2018} dialogues based on users' leisure time activities. The aim of these dialogues is to explore users' leisure time activities, and if necessary, nudge them towards a more active lifestyle.
		\item \textbf{Nutrition} \cite{Sayas_nutrition2018} dialogues focused on users' nutritional habits. The goal of these dialogues is to explore the nutritional routines of the users and, if necessary, try to make the users vary those routines in order to form nutritional habits that are potentially more healthy.
	\end{itemize}

%For future work, the already available dialogues will be improved and new ones will be implemented (e.g. family relationships). Also, the use of an alternative DM based on statistical approaches is planned.

\subsection{Natural Language Generation}
The Natural Language Generation (NLG) component maps the abstract dialogue acts provided by the DM to natural language constructs (in Spanish, French or Norwegian), written in orthographic form. So far, the NLG component has been developed from a reduced database of coaching turns. The data have been extracted from video recordings of real user sessions with a professional coach, and some handmade dialogues created by a professional coach. In the process of labeling, two types of labels have been used: (1) based on the GROW coaching model~\cite{alexander2010behavioural}, and (2) based on linguistic features needed to construct the text. Considering the restrictions in the amount of training data, the  NLG currently represents a rule-based system, using a template-based approach~\cite{oh2000stochastic}. Future work, however, aims to design a new NLG component based on a seq2seq neural network~\cite{duvsek2015training} (note: this plan of implementing a seq2seq neural model only concerns the NLG part of the EMPATHIC VC and  will not affect the other components).

\subsection{Text-to-Speech Synthesis}
A Text-To-Speech (TTS) component converts any text into a spoken message. For EMPATHIC, TTS speaking styles should be fully compatible with the role of the VC communicating with elderly users. The Acapela TTS system employs a range of internally developed technologies, such as unit selection or parametric synthesizers based on Hidden Markov Models (HMM's) or Deep Neural Networks (DNN's). During the project, all of these technologies are adapted for the EMPATHIC communication and compared during evaluation with end-users. Professional speakers will be recorded to capture the communicative role of the VC with the aim to reflect the expressive possibilities of the dialogue system by coherent audio responses to the user's emotional state so as to support the credibility, naturalness and adaptability of the full dialogue chain. So far, Acapela already recorded a Spanish professional speaker enacting a VC for the elderly and trained a snit selection, and initial DNN synthesis systems on this corpus. We run evaluations of this elderly coach style TTS in terms of naturalness and intelligibility. A Spanish coaching TTS voice is already available and integrated in the mid-term version of the EMPATHIC VC. Upon the evaluation of the Spanish system, Acapela will fine-tune the process and develop the French and Norwegian voices.

\subsection{Emotional Agents} \label{sec:agent}
\begin{figure}[!t]
 \centering
 \includegraphics[width=\linewidth]{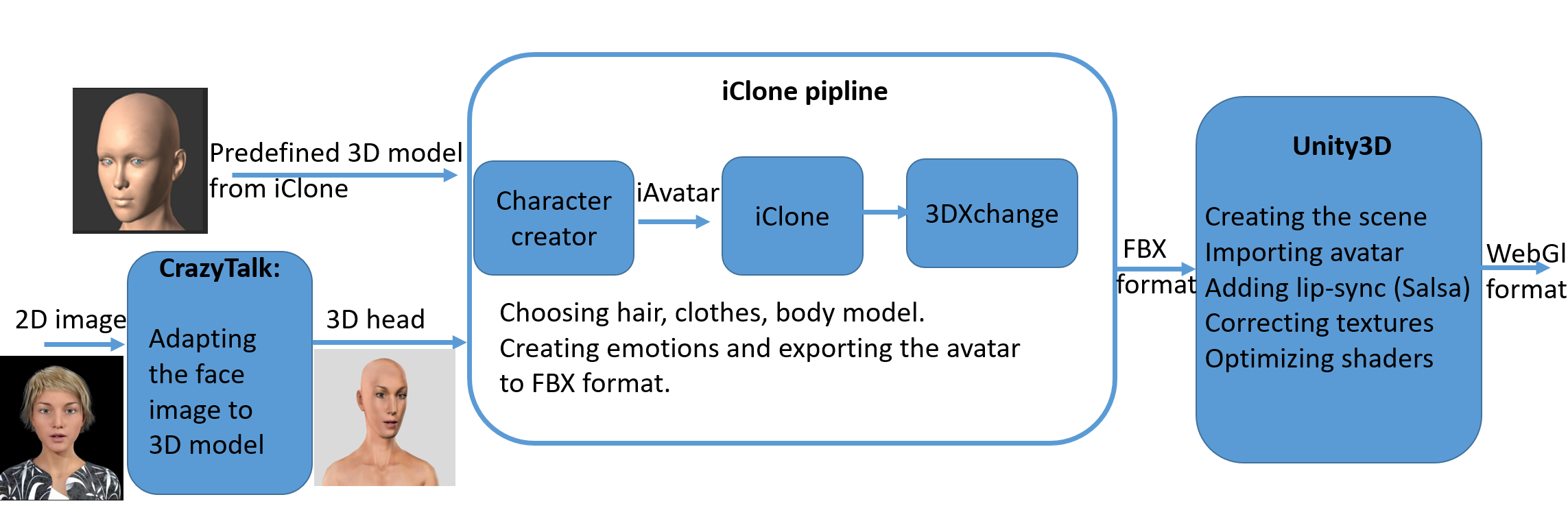}
 \caption{Workflow to generate 3D virtual coaches.}
 \Description{Workflow to generate 3D virtual coaches}
 \label{fig:WorkflowVC}
\end{figure}

For initial prototyping purposes (cf. Section~\ref{sec:woz}), we used a multi-step creation process (cf. Figure~\ref{fig:WorkflowVC}) to build five virtual agent coaches (3 female and 2 male) named Natalie, Alice, Lena, Christian and Adam. The first of the steps depended on the origin of the 3D model. The coaches Alice and Adam  were created based on 2D images. For this, we had to create their 3D model from a 2D image using CrazyTalk\footnote{https://www.reallusion.com/crazytalk/} and then exported this to the RLhead format. For the coaches Christian, Natalie and Lena, who were created based on 3D models predefined by iClone\footnote{https://www.reallusion.com/iclone/}, we skipped the CrazyTalk step. Second, we imported the RLhead to the Character creator. This tool helped us create realistic-looking 3D human models. We fixed the 3D model design, imported 3D clothing designs and generated humanoid animations with extensive customization tools. After that, a model was imported into iClone. At this step, we used iClone to blend character creation, animation and scene design into a real-time engine, and to edit them in 3DXchange. Next we exported the model and animation using 3Dxchange into the FBX format. Then, we imported the FBX file as a new resource into Unity3D\footnote{https://unity3d.com/fr}. The transition from iClone to unity degrades the appearance of the model. To solve this, we had to correct the texture and optimize the shader of the model. Still in this step, we added the lip-synchronization using the SALSA plugin. The audio was processed so as to automate four basic mouth positions, which are the basis for lip-sync approximation. Instead of pre-processing or mapping shapes to audio markers, mouth movements were procedurally applied to a minimal set of mouth shapes to provide variation. It uses a combination of waveform analysis and four mouth shapes to produce high-quality lipsync approximation. The last step was to build the WebGL format within the Unity environment.

% - SYSTEM INTEGRATION %%%%%%%%%%%%%%%%%%%%%%%%%%%%%%%%%%%%%%

\section{Status: Technology Integration}
\label{sec:technology}
To integrate the different software components of the EMPATHIC-VC while meeting the challenges of a low-latency interactive system with the additional security, confidentiality and privacy requirements for health information, we took a multi-stage approach. First we defined up front a model for development which uses fully separated containers for each component, with communication between components over sockets and messaging via a global message queue. Next, a review process was conducted for each of the components, covering the component container layout and requirements, capabilities and testing approach.
Subsequently, a dedicated integration environment was set up with the supporting tools such as the container orchestration and message queue. Finally, each component was tested independently in the integration environment, both initial smoke testing and then testing component inputs and outputs.

%To support these modules we have selected several open source packages to provide infrastructure, including Linux operating system, Docker container, Kubernates orchestration, Wordpress CMS, Apache http server, MySQL, Apache ActiveMQ, OpenVidu and Kurento.

Once all components are validated in the integration environment, according to the above described procedure, we will be ready to test the entire system. For the purpose of testing we have split the system into four sub-systems:
\begin{itemize}
\item An ``inbound'' system, using pre-recorded input, joining ASR, NLU and DM.
\item An ``outbound'' system, using pre-generated DM output, joining NLG, TTS and the virtual agents.
\item A ``user interaction'' system, joining the Web UI, Web A/V proxy and also test recordings.
\item A ``human sensing'' system, joining the emotion detection for speech, text, face and gaze, and the biometric authentication.
\end{itemize}
%On completion of the sub-system tests we have planned a series of end-to-end tests with expected outcomes.  This are expected to be completed before the mid-term review in May 2019.
In addition to the integration of the VC components we started to work on the provision of secure cloud connectivity. For this we have designed a review process covering:
\begin{itemize}
\item the secure connection over WebRTC between participant devices and the EMPATHIC-VC system;
\item the secure configuration of host servers;
\item secure remote administration;
\item secure software development practices;
\item secure storage and encryption of participant data;
\item and physical security at field trial hosting sites. %(validated at Bilbao and London; Paris and Oslo to be completed by August 2019)
\end{itemize}

% - HUMAN-SYSTEM INTERACTION %%%%%%%%%%%%%%%%%%%%%%%%%%%%%%%%%%%%%%

\section{Status: User Interaction}
\label{sec:interaction}
%Defining and analyzing the needs and wishes of the end-user group, early on the development process, it is a key aspect for the final solution's success. 
In order to ensure acceptance of the EMPATHIC VC, we have to show an added value for the end-user. The creation of this value proposition is an iterative and contrasted process that involves interdisciplinary collaboration~\cite{pagliari2007design}. To start this, we analyzed relevant target groups in Spain, France and Norway. From those studies we were able to extract general as well as country-specific end-user traits, leading to an archetypical end-user definition with some specific factors that may vary from country to country. 

\subsection{Defining the Target Population}
As a common definition, we can consider our target population as ``Young Olds'', aged 65 to 79 years, with a healthy and active life, characterized by the continuation of their former lifestyle after retirement, yet focused more on enjoyment and leisure activities.
%In this phase, there is a change and reorganization of the new leisure time for which they did not previously have enough time, and that make them feel good or needed. The main focus is enjoyment and activity.
It is important to highlight the two main indicators used for this definition. The first indicator concerns the healthy life years at birth; this indicator shows the average of life years sans diseases.
%the definition of these ``Young Olds'' population, as per Eurostat\footnote{https://ec.europa.eu/eurostat/data/database}. Since the life expectancy estimation does not consider factors such health and quality of life. 
%This is an objective indicator of the population health, based on records and healthy census. 
The second indicator concerns the healthy life expectancy based on self-perceived health; this indicator shows the population's self-perception of health. Thus, it include factors such as economic status, emotional problems and social relations. This is a subjective indicator based on surveys, perception records and/or self-assessments\footnote{https://ec.europa.eu/eurostat/data/database}.

Therefore, the term ``Young Olds'' refers to people older than 65, who perceive their health as good or very good. Yet although, their self-perception is good, ``Young Olds'' may already have some type of disease or sickness\footnote{Encuesta Europea de Salud en Espana 2017, Pub. Pub. Instituto Nacional de estad\'{i}stica (INE)}. 

%Based on Eurostat indicator, population from 79 years old onwards loses this self-perception of good health status\footnote{Encuesta Nacional de Salud 2017. Pub. MSCBS-INE}. 

\subsection{Understanding the ``Young Olds''}
Before describing the priorities defined by our analysis, it is worth to highlight four basic recommendations to promote user acceptance of ICT by seniors, defined by the European Active and Assisted Living (AAL) programme. Those recommendations are \footnote{http://www.aal-europe.eu/wp-content/uploads/2015/02/AALA\_Knowledge-Base\_YOUSE\_online.pdf}: 

\begin{itemize}
    \item Provide clear additional value and benefit of the solution;
	\item Balance between supporting the users and activating them;
	\item Maintain simplicity on the interaction user-solution;
	\item Provide joyful experiences;
\end{itemize}

Those recommendations act as a starting point for integrating priorities into a solution design. 

Regarding our analysis, family has been revealed as the main priority across different countries. On average, more than two thirds of seniors with children see them several times per month. This ratio is even higher when children live close by. In addition, the use of ICT tools is significantly higher when it is used for communication purposes with family members (especially children and grandchildren). Therefore, the involvement of family members as part of the solution will strongly increase the acceptance and usability of the solution. 

Another key point to promote the ICT use among elderly is to show the clear benefit of a solution. For example, seniors are willing to learn to use ICT tools if this provides more interaction with their younger family members. 

As mentioned before, more than half of our target population perceives their health as ``good'' or ``very good'', although a common fear is the physical decline associate to ageing. This could be one strong reason why, independent of the country, more than half of the people we talked to perform some type of physical activity on a weekly basis (note: most commonly walking). Supporting those activities, providing motivational strategies, gamification tools and/or professional advice may thus be seen an opportunity to increase the acceptability of ICT. Even better, it would not only support and enrich a leisure activity that is already common among the target population, but also promote ``healthy habits'' and ``well-being''.

Following a similar approach, nutrition is an area that combines ``joyful'' experiences and ``healthy habits''. Cooking, planning meals, going out to restaurants, increasing expenditure on food etc. are related activities that are more promoted with representatives of the target population than with younger people. In that sense, changes in the nutrition habits can be seen as an indicator for physical or physiological changes. On the other hand, promoting and motivating a healthier nutrition habit can positively influence a human's physical and emotional status.

%The target population, with independence of the country of origin, spend a significantly amount of time watching TV in a daily basis. Therefore, this fact should be considered on the solution development, based on a strong habit already hold by the targeted end-users. Finally, another leisure activity common on more than half of the senior population among the three countries is traveling. Information, advises, recommendation or just conversations based on that topic could support the engagement and activation of the end-users.
%In the other side, one relevant difference among target populations, 
While motivational factors and physical as well nutritional habits were similar with our target populations in Spain, France and Norway, we did also find differences. Those mainly relate to the familiarity and use of the Internet. Here it was shown that Seniors from Norway have a significantly higher percentage of people that use the Internet on a regular basis, in particularly when compared to Spain (note: France lies in between). This information is relevant, as personal experience with technology is a relevant factor influencing technology acceptance. 
%Therefore, it would be expected that Scandinavian populations would have higher scoring on the technology acceptance than Spanish one. Thus, this factor should be considered for technology acceptance evaluation among countries.

\subsection{Aspects of User Acceptance} \label{sec:acceptance}
One aspect to focus on in human-agent interaction is the user's level of technology acceptance. The concept was introduced by Davis~\cite{davis1989perceived} in the attempt to explain people's acceptance (or not acceptance) of an interactive system. It led to the development of the Technology Acceptance Model (TAM), a questionnaire where acceptance is assessed in terms of a user's perceived usefulness, and perceived ease of use~\cite{davis1989perceived} of the system. TAM was extended into TAM2 in 2000~\cite{venkatesh2000theoretical}, adding two theoretical constructs that accounted for a user's social influence and for how well a user's work goals are supported by the interactive system. TAM2 evolved into the Unified Theory of Acceptance and Use of Technology (UTAUT)~\cite{venkatesh2003user} and later into UTAUT2, where hedonic motivations (the fun or pleasure derived from using a technology), price values (trade-off between perceived benefits and monetary costs), and habits were added to the original questionnaire, as further determinants theorized to affect user's behavioral intentions and use behavior~\cite{venkatesh2012consumer}. Finally, the Almere questionnaire was developed as a further evolvement of UTAUT2 objecting that the latter was developed without accounting for variables that relate to social interaction with robots or virtual agents and without considering seniors as potential users~\cite{tsiourti2014virtual, heerink2010assessing}. Following the same reasoning, Hassenzahl developed the AttrakDiff  questionnaire\footnote{AttrakDiff(tm)Internet Resource -- http://www.attrakdiff.de.}~\cite{hassenzahl2018thing, hassenzahl2004interplay}, a four cluster test, where each cluster, composed of 7 items, was assessing a desired user requirement. 

It must be noted that, currently, all the theoretical formulations of questionnaires aiming to assess the user experience of an interactive system are, to a certain extent, dated, since those systems are increasingly more complex, showing humanoid appearance, and human features. Thus, even though the theory for defining user experience is still valid, new concepts have to be accounted for, in order to have a fair assessment of modern interactive systems. This is why, to date, there are no systematic investigations devoted to assessing the role of virtual agents' features exploiting the above mentioned questionnaires. Furthermore, seniors have only been involved in a very limited number of studies on virtual agent's. In the few they have been involved in, it has been shown that they clearly enjoy interacting with a speaking synthetic voice produced by a static female agent (note: these were 65+ aged seniors in good health~\cite{cordasco2014assessing}), and that such seniors are less enthusiastic than impaired people in recognizing the agent's usefulness~\cite{yaghoubzadeh2013virtual}. The only comparison among user's age we know about, was conducted by Stra\ss  mann \& Kr\"{a}mer~\cite{strassmann2017categorization}. The study was \textit{``a qualitative interview study with five seniors and six students''} and showed that senior users prefer embodied human like agents over machine or animal-like ones. No information were obtained on the gender of the agents as well as their pragmatic and hedonic features, as advocated by the TAM, UTAUT, and AttrakDiff questionnaires discussed above.

Therefore, the Empathic project aims to \textit{``develop causal models of [agent] coach-user interactional exchanges that engage elders in emotionally believable interactions keeping off loneliness, sustaining health status, enhancing quality of life and simplifying access to future telecare services''}. To this end, an initial research step was the development of an ad-hoc questionnaire to assess the pragmatic and hedonic features of the to be developed EMPATHIC VC. The questionnaire was developed through an iterative process that involved several experiments and the exploitation of the theoretical concepts already advocated by the authors of TAM 2, UTAUT2, and AttrakDiff, and the inclusion of new theoretical considerations regarding our users' age. The goal of the questionnaire was also to provide information on the user's preferences regarding the agent's physical and social features, including its face, voice, hairdo, age, gender, eyes, dressing mode, attractiveness, and personality.
%and these features must be considered together with the age of the users which are seniors aged 65+ years. 

The first experiment with the aim to build and evaluate such a questionnaire (and start collecting said data) was conducted in Italy and involved 45 healthy seniors (50\% female), aged 65+~\cite{esposito2018seniors}. As this study was conducted before any of the agents described in Section~\ref{sec:agent} were built, our stimuli were based upon the four conversational agents proposed by the Semaine project\footnote{http://www.semaine-project.eu/}, each possessing different personality features able to arise user specific emotional states i.e., Poppy (female, expressing optimism), Obadiah (male, expressing pessimism), Spike (male, expressing aggression) and Prudence (female, expressing a high degree of pragmatism)~\cite{ochs2010virtual}. For each agent, a video-clip was extracted from the videos available on the Semaine website. In order to contextualize them to the Italian culture they were renamed, using names very popular in the local area (i.e. Serena, Gerado, Pasquale, and Francesca). Agent's names and video durations were carefully assessed by four people. The final set of stimuli consisted of 4 video clips, each 10 secs long showing the agent's half torso, all of the same dimensions, acting as if they were speaking while the audio was mute.

%In particular the selected agents were:
%\begin{itemize}
%\item Poppy (a female agent, renamed Serena), committed to expressing optimism
%\item Obadiah (a male agent, renamed Gerardo),  deputed to express pessimism
%\item Spike (a male agent, renamed Pasquale),  deputed to express aggression
%\item Prudence (a female agent, renamed Francesca), committed to express a high degree of pragmatism. 
%\end{itemize}

The preferences the target group had towards each of the proposed agents were assessed through a first version of our questionnaire, structured in 4 clusters each containing 7 items devoted to assess the practicality, pleasure, feelings, and attractiveness experienced by participants while watching the agent video-clips. The items proposed in each cluster exploited the theoretical foundation inherent to the UTAUT2, Almere, and AttrakDiff questionnaires. Results showed seniors' positive tendency to initiate an interaction with the agent, with a strong preference towards agents with a positive personality. That is, Francesca and Serena scored always significantly higher than Pasquale and Gerardo for the pragmatic, hedonic and attractiveness features. Although seniors had not been informed of the agents' personality, somehow they had perceived  negativity or positivity from the dynamics of their facial expressions, which triggered a behavior of acceptance/rejection, suggesting that participants have preferences for positive facial dynamics. These results, however, required deeper investigations as all agents showing positive facial dynamics were females, hinting towards a potential gender influence on the processing of emotional facial expressions (cf.~\cite{marsh2005effects, rotteveel2004automatic}). Furthermore it must be noted that in this first study agents were dynamically moving their lips as if they would be speaking. Their voice was, however, muted which might have had an effect on people's perceptions. 

These aspects were accounted for in a subsequent experiment, conducted again in Italy~\cite{esposito2018bseniors}, but this time with agents created with BOTLIBRE\footnote{http://www.botlibre.com}. 
%which that allow users to freely create a customer service virtual agent according to their preferences and goals, providing a wide set of agents with different humanoid semblances. The selection of the agents was made by three experts on the basis of preferences dictated by the agent's professional and non-emotional appearance. The selected four virtual agents two males and two females, received 100\% of agreements among the experts. 
Also these agents, which were selected by 3 experts, showed half their torso, with definite clothing. Again, so as to contextualize, we named them in accordance with typical names for the region, i.e. Michele, Edoardo, Giulia and Clara. Each agent was provided with a different synthetic voice, producing the following sentence pronounced in Italian: \textit{``Hi, my name is Clara / Edoardo / Giulia / Michele. If you want, I would like to assist you with your daily activities!''}. The synthetic voice was created through the website Natural Reader\footnote{http://www.naturalreaders.com}. The voices (recorded using the software Audacity\footnote{https://www.audacityteam.org/}) were embedded into each agent's video-clip which had an average duration of approx. 6 seconds. The proposed agents did not show a particular personality. Care was taken to ensure that no emotion was depicted by their faces and they were shown on the same background, avoiding exaggerated cloths colours or facial features. This second experiment, again devoted to assessing senior's preferences towards each of the proposed agents, exploited a new version of our questionnaire. It had 6 clusters:

\begin{itemize}
\item \textbf{Cluster 1:} 10 items focusing on the usefulness, usability, and accomplishment of the tasks of the proposed system, i.e. the system's pragmatic qualities (PQ). High scores in the PQ dimensions indicate that the users perceive the systems as well structured, clear, controllable, efficient and practical. 
\item \textbf{Cluster 2:} 10 items focusing on motivations, i.e. the reason why a user should own and use such an interactive system, i.e., the system's stimulating hedonic qualities (HQS). A system receiving high scores in the HQS dimensions is meant to be original, creative, captivating.
\item \textbf{Cluster 3:} 10 items focusing on how captivating, and, of good taste the system appears, i.e. the system's hedonic qualities of feeling (HQF). A system receiving high scores in the HQF dimensions, is considered presentable, professional, of good taste, and bringing users close to each other.
\item \textbf{Cluster 4:} 10 items on the subjective perception of the system's attractiveness (ATT), which is the hedonic dimension that gives rise to behaviors as increased use, or dissent, as well as, emotions as happiness, engagement, or frustration. 
\item \textbf{Cluster 5:} 4 items assessing the type of professions seniors would endorse to the proposed agents, among which were welfare, housework, security, and front desk jobs.
\item \textbf{Cluster 6:} 3 items assessing agent's age range preferences. 
\end{itemize}

Each questionnaire item required a response given on a 5-point Likert scale ranging from 1=strongly agree to 5=strongly disagree (3=I don't know). Since all clusters contained positive and negative items evaluated on a 5-point Likert scale, scores from negative items were corrected in a reverse way. This implies that low scores summon to positive evaluations, whereas high scores to negative ones. Each participant was first asked to provide answers to the items related to demographic information and  user technology savviness, then they were asked to watch each agent's video-clip and immediately after to complete the items from the remaining 6 clusters. 

\paragraph{Final Results}
Our results show that seniors clearly expressed their preference to interact with the female rather than the male agents. This was true for  the pragmatic, hedonic stimulation, hedonic feeling and the attractiveness dimensions, and independent from their gender and technology savviness (note: between the two proposed female agents, senior's preferences for Giulia scored statistically significantly higher than those attributed to Clara). Consequently, the data suggests, that seniors' willingness to be assisted by a virtual agent is strongly affected by the gender of the proposed agent, and that, up to now, female agents seem to be the preferred choice. In addition, these two preliminary experiments allowed the definition of a questionnaire on virtual agent acceptance, which also contains a demographic information section and a section on user technology savviness. We call this the Virtual Agent Acceptance Questionnaire (VAAQ). So far, it has been translated from English into French, Norwegian, Spanish, Italian, and German. 

\subsection{A Simulated Virtual Coach} \label{sec:woz}
%In order to simulate various features of EMPATHIC-VC (in particular those, which require ongoing development and improvement such as speech recognition and language understanding as well as user experience and acceptance), the goal was to build and consequently use a simulated Wizard of Oz (WOZ) component.
Various modules of the EMPATHIC VC will require ongoing development and improvement, e.g. to advance speech recognition and language understanding, or foster user experience and acceptance. In order to simulate such modules while they are being developed, the goal was to build and consequently use a simulated Wizard of Oz (WOZ) component. WOZ constitutes a prototyping method that uses a human operator (i.e., the so-called wizard) to simulate non- or only partly-existing system functions. In language-based interaction scenarios, like the ones envisioned by EMPATHIC, WOZ is usually used to explore user responses and the consequent handling of the dialogue, to test different dialogue strategies or simply to collect language resources (i.e., corpora) needed to train technology components. In EMPATHIC, however, the goal is to use WOZ beyond this traditional prototyping stage, and make it a fallback safety net for situations in which the automated coach may be unable to respond. That is, the goal is to develop a system component, which initially serves as a prototyping tool supporting the research on language-based interaction and dialogue policies, but then becomes an always-on backup channel dealing with those user requests the system is incapable of handling by itself. A first version of this tool has been built and consequently used in several user studies. 

\subsubsection{Technical Setup}
As the specification and integration of the the EMPATHIC VC is still ongoing (cf. Section~\ref{sec:technology}), yet user feedback is urgently needed to inform the design of technology components, we decided to work on two separate WOZ systems. The first one, referred to as the EMPATHIC WOZ Platform, acts as a stand-alone tool, which is being developed independently from the EMPATHIC VC architecture, despite being a testbed for feasibility evaluations regarding the different technologies to be used. In doing so, we were able to perform initial user studies before final decisions on the EMPATHIC architecture were taken. On the contrary, the second WOZ system, referred to as the EMPATHIC WOZ Component, is meant to become a component of the final EMPATHIC coach for which it is implemented and adapted in accordance with the overall EMPATHIC system architecture. This component is currently in development and thus not yet used for user studies.

\subsubsection{The EMPATHIC WOZ Platform}
Several researchers have worked on WOZ tools before (e.g., \cite{munteanu2000mdwoz, fiedler2002supporting, hundhausen2007woz, davis2007sketchwizard, smeddinck2010quickwoz, lu2011polonius, villano2011domer}), but only few of those tools are openly available for implementation and adaption. One such tool is the WebWOZ Wizard of Oz Prototyping Platform~\cite{schlogl2013webwoz}\footnote{https://github.com/stephanschloegl/WebWOZ}, which has already been employed by a number of previous projects (e.g., vAssist~\cite{schlogl2014designing}, Roberta Ironside~\cite{lee2017first}). Building upon the experience of these projects, we used the platform as a core system and implemented the following EMPATHIC-specific adaptations:

\begin{itemize}

\item \textbf{Audio Transmission and Recording:} 
Previous versions of WebWOZ required a separate audio channel to transfer a study participant's voice input to the wizard. Usually, such was achieved using some sort of Voice-over-IP tool such as Skype or Google Talk. To avoid 3\textsuperscript{rd} party tool usage we directly integrated audio transmission between a study participant and a wizard via a connection channel based on the WebRTC standard. 
%WebRTC is a free, open project supported by the W3C and many technology providers, providing browsers and mobile applications with Real-Time Communications (RTC) capabilities via simple APIs. 
In our adapted WebWOZ Platform, WebRTC does not only serve as a communication channel but also handles the recording of sessions, allowing for the collection of relevant language resources needed to inform the design of future EMPATHIC language components (e.g., ASR, NLU, DM, etc.). In addition, the technology will be used as a core technology for the final EMPATHIC VC architecture, for which its implementation into the WebWOZ Platform acted as a test case evaluating its feasibility.

\item \textbf{Video Transmission and Recording:} 
In addition to the audio transmission channel described above, we integrated video transmission and recording between participant and wizard. Also here, WebRTC acted as the core technology. The video link, which was added to the wizard interface, allows the wizard to see the participant's face, providing important contextual information and thus supporting the cognitively rather demanding task of simulating machine behaviour (cf.  Figure~\ref{fig:wizardinterface}). 

\begin{figure}[!t]
  \centering
  \includegraphics[width=\linewidth]{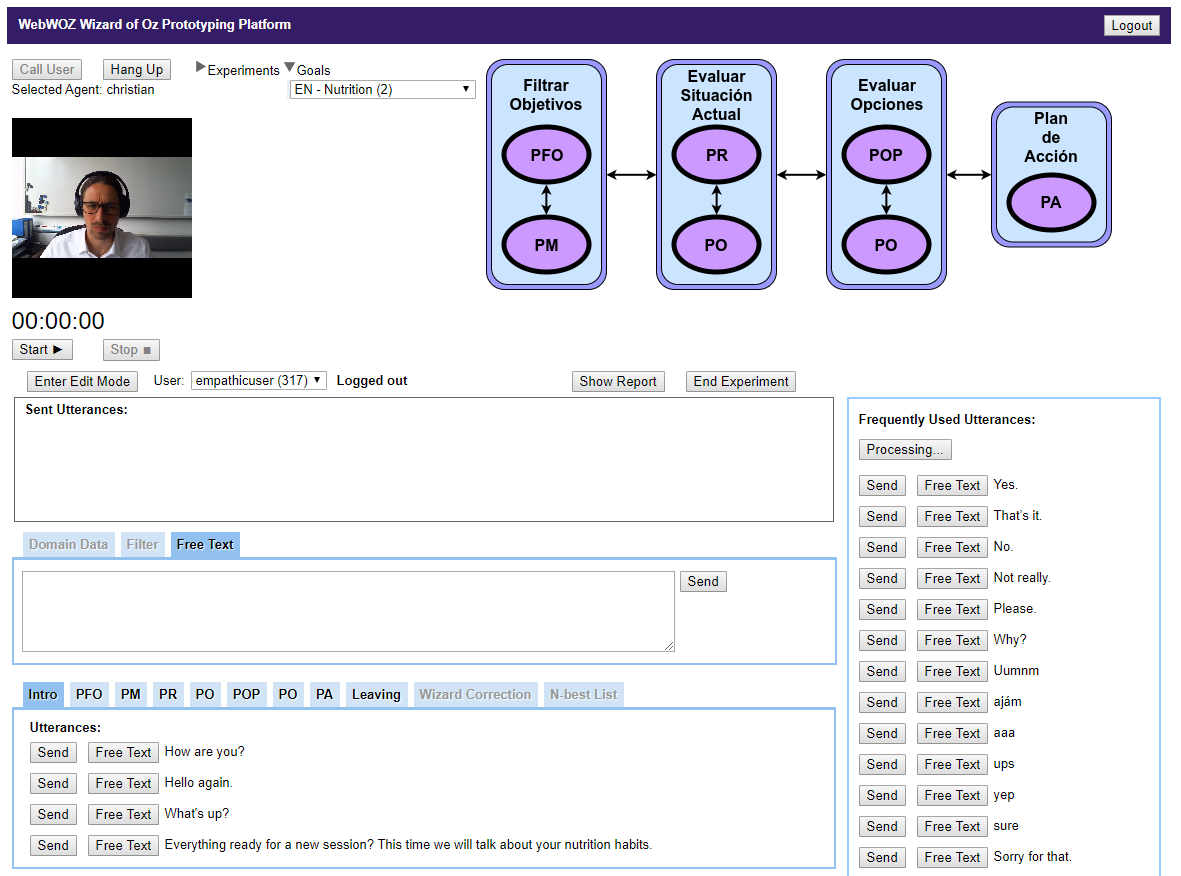}
  \caption{EMPATHIC WOZ Platform wizard interface incl. live video feed and flow-chart graph.}
  \Description{EMPATHIC WOZ Platform wizard interface}
  \label{fig:wizardinterface}
\end{figure}

\item \textbf{Flow-chart in Wizard Interface:}
In order to help the wizard follow a systematic interaction path, we further added a flow-chart graph to the wizard interface (see Figure \ref{fig:wizardinterface}). This graph shows the optimal flow of dialogue steps and thus acts as an interaction manual supporting the selection of pre-defined utterances to be sent to a study participant. 

\item \textbf{Web-based Scenario Upload Mechanism:}
The definition of pre-defined utterances to be sent to a study participant counts as a key feature of a language-based WOZ tool. In order to speed up the definition of these utterances and at the same time allow developers and interaction designers to work with standard tools, we integrated an Excel (.csv) import feature. A similar feature was already available in earlier implementations of WebWOZ \cite{schlogl2014designing}, yet such needed root access to the server infrastructure where the platform was hosted. We extended this feature by a simple web-based upload mechanism, so that respective user rights are no longer required. 

\item \textbf{Agent Interface incl. Text-to-Speech Synthesis:}
The client interface of the original WebWOZ Prototyping Platform does not offer any agent or avatar feature. Hence, in order to obtain feedback on our EMPATHIC virtual agent designs, we implemented five different agent prototypes (3 female, 2 male) and connected them to the wizard interface (cf. Section~\ref{sec:agent}). All agents use similar core features (size, background, facial features, etc.) and integrate, depending on the environment setup, Spanish, French and Norwegian (as well as German, Italian and English) text-to-speech synthesis provided by Acapela (note: female and male agents use different voices).
\end{itemize}

With this setup, a human wizard is currently able to remotely control a virtual agent in any of these languages. Lessons learned from first tests using this setup are reported next.

%the EMPATHIC consortium member
\begin{figure}[!t]
  \centering
  \includegraphics[width=\linewidth]{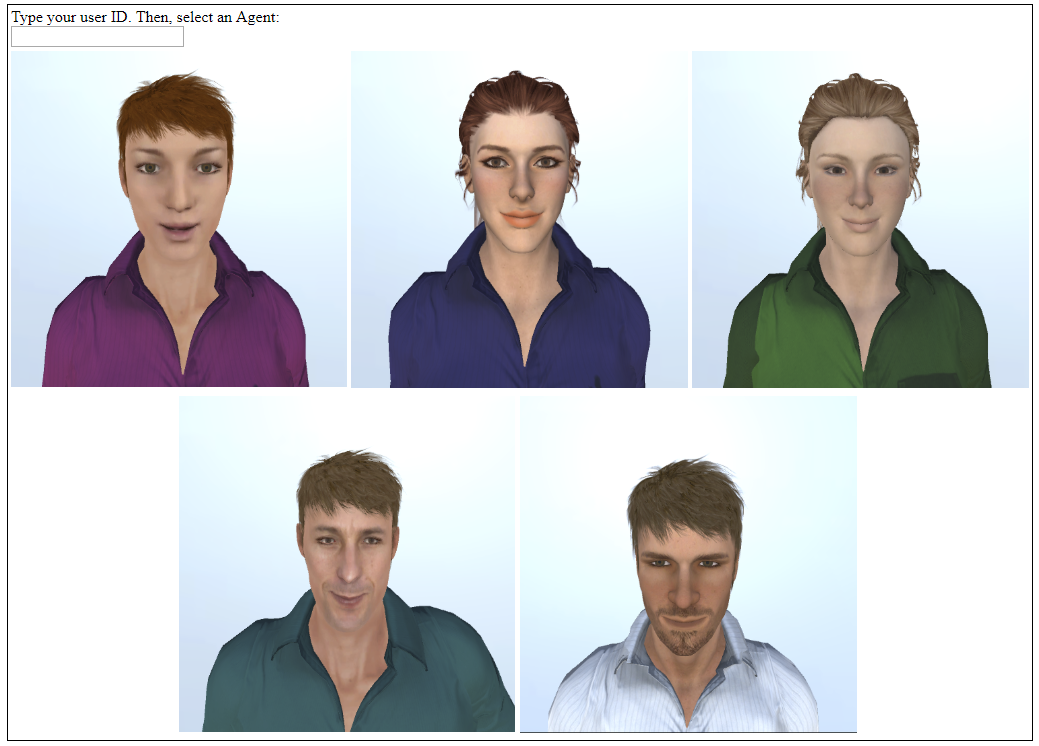}
  \caption{EMPATHIC WOZ Platform client interface featuring five different agents to interact with.}
  \Description{EMPATHIC WOZ Platform client interface}
  \label{fig:clientinterface}
\end{figure}

\subsection{Lessons Learned from User Studies}
A total of 176 WOZ user studies (\`{a} 2 sessions) have so far been conducted (i.e., 68 in Spanish, 54 in French, and 54 in Norwegian). The following insights are based on feedback provided by wizards, who simulated the EMPATHIC VC, as well as study participants.

\subsubsection{Insights Regarding the Study Setup}
Experience has shown that at least two people are required to realistically conduct a WOZ user study -- one who acts as a human simulator, i.e. the wizard (usually sitting in a different location), and one who acts as a facilitator, greeting study participants, introducing them to the study purpose, administrating questionnaires, and helping the participants in cases of confusion. A setup without facilitator, i.e. without a second person, did in our case not seem feasible. 
From a procedural point of view, we further found it imperative that, once the interaction with the simulated agent has started, the facilitator needs to leave the room, because otherwise the participant tends to look at and talk to the facilitator instead of conversing with the actual agent. This behaviour may be explained by a participant's lack of reassurance when interacting with a novel technology. Taking the additional person out of the room helps eliminate this potential distraction, yet it may also increase a participant's level of anxiety. 

\subsubsection{Insights Concerning Study Participants}
In general, we found that the concept of a virtual agent seems rather frightening to many people of the targeted age group (i.e. aged 65 or older). While we did use face-to-face meetings to overcome this fear as much as possible, it should be noted that for this type of technology anxiety poses a significant challenge, particularly when it comes to the recruitment of study participants. Consequently, recruitment via flyers/posters was difficult (even when conducted in senior centres or elderly homes). However, we found that recommendations coming from other participants who had already taken part and enjoyed the study, helped mitigate the problem. Still, a lot of personal coaching was usually required to make people feel comfortable. Here, our experience has shown that participants needed approx. 10 minutes to `lose their fear' regarding the technology -- in particular, when studies took place somewhere away from peoples' homes or familiar living environments. A technical setup, which would allow studies to be mobile and, thus, be brought to potential study participants, may therefore help circumvent some of this felt uneasiness. 
With respect to the study inclusion criteria, the studies have shown that elderly people are rather pessimistic when evaluating their personal health status. That is, while initially we were searching for `healthy' participants aged 65 or older, we had to realize that most representatives of this group would not include themselves due to minor health issues they perceived preclusive (e.g. minor hearing problems). A slight change in wording has helped tackle this problem. What remained difficult, however, was the recruitment of people with depression. 
As for the interaction, it seemed important that participants thought they would interact with a prototypical system. This helped keep the expectations regarding speed and accuracy low. In this context, the speed with which a simulated system responds may be seen a particular challenge. Especially in cases where the wizard could not use a pre-defined utterance and, thus, had to type a response. An additional challenge with this generation of on-the-fly utterances concerns the great potential for typo's and other mistakes, which are forwarded to the TTS and, consequently, spoken out loud to a study participant. However, being aware of the prototypical status of the system, study participants were rather tolerant towards these types of issues.

\subsubsection{Insights Concerning the Dialogue}
With respect to the scenarios, participants were usually pre-informed about some of the content to be addressed by the coach so that they could think about relevant topics in advance (e.g. they were told to think about certain health goals they would like to achieve before starting the conversation). Such was necessary to keep the interaction going and reduce the number of ``yes/no'' answers. Still, in particular with respect to the the nutrition scenario, it was difficult to keep the conversation flowing, as the scenario was looking for personal goals, yet people were often satisfied with their status-quo and, thus, did not find much to talk about. This somewhat restricted the number of available interaction turns, for which we had to slightly shift participants' foci to other nutrition-relevant topics so as to keep the conversation alive. To this end, the pre-defined utterances that were prepared for the wizard seemed rather limited in scope as well as in variation, which significantly increased the use of the (arguably much slower and more error-prone) free-text feature of the platform. A way of making the conversation more flexible, may be found in a speech input interface for the wizard. Yet such a feature, while in consideration for the EMPATHIC WOZ Component, has so far not been integrated. 
Changing the conversational focus due to missing participant goals also caused some side effects. Particularly in France, participants often felt insufficiently `coached'; i.e. they had the impression that the virtual coach wanted their information, yet did not provide them with any advice on what they should do in order to change their habits. 
Finally, from a conversational point of view, we found that different types of back-channelling (i.e. approving a participant's input) had a significant influence on the `smoothness' of the conversation. That is, while rather basic approval utterances such as ``interesting'' or ``good'' seemed to distort the conversation, other strategies which re-used participants' words or sentence structures (e.g. Participant: ``I like to walk 2 hours every day''; Agent: ``You walk 2 hours every day?'') helped in keeping participants engaged and consequently the conversation flowing. 

\subsubsection{Insights Regarding Administered Questionnaires}
Regarding the study closure and debriefing phase, study participants perceived the number and length of the administered questionnaires as too much (note: this also included our own questionnaire whose development was presented in Section~\ref{sec:acceptance}). In addition, some of the questions were hard to understand or potentially ambiguous and, thus, study participants found them to be rather difficult to answer (note: this led to a high number of ``I don't know'' answers). In particular, questions concerning more abstract concepts or terms as well as feelings not usually connected to a human-agent dialogue (e.g. hedonic features) seemed to pose problems. Often it was the wording, which played a significant role here and might, thus, be adapted in future studies. Finally, unexpectedly high depression scores found with healthy study participants caused some doubts regarding the used scales, which will require further investigation.

\subsubsection{Insights Regarding Technical Issues}
From a technical point of view, a re-occurring problem seemed to be connected to the session management, which influenced the connection between the wizard and respective client. The issue is currently under investigation, with a viable solution expected to be implemented in the upcoming weeks. A second challenge concerned the technical support during user studies. Given that the entire EMPATHIC WOZ Platform is currently hosted in a location (i.e. the UK) different from where studies are conducted, and technical support provided from yet another location (i.e. Spain), technical problems often caused long delays. Local setups incl. respective technical support -- as it is planned for the final EMPATHIC VC -- may thus be considered in future studies. 

% - CONCLUSION AND ACKOWLEDGEMENTS %%%%%%%%%%%%%%%%%%%%%%%%%%%%%%%%%%%%%%

\section{Conclusion}
\label{sec:summary}
In this paper we reported on the mid-term achievements of the H2020 EMPATHIC (\textit{Empathic, Expressive, Advanced Virtual Coach to Improve Independent Healthy-Life-Years of the Elderly}) project. Those achievements include, on the one hand, significant efforts put into the development and integration of various technical components required to run a modern, virtual agent based dialogue system geared towards supporting elderly people in their daily activities; on the other hand,  a number of user studies aimed at understanding said user group (i.e., healthy seniors aged 65+) and their preferences with respect to agent acceptance. Results of these efforts are manifested in a working WOZ prototyping tool for simulating human-agent interaction, a multi-lingual (i.e. Spanish, Norwegian, French, Italian, German and English) questionnaire assessing virtual agent acceptance, and a better understanding of particular technical challenges inherent to the provision of a web-based, secure, responsive and reliable agent platform.

\section{Acknowledgments}
The research presented in this paper is conducted as part of the project EMPATHIC that has received funding from the European Union's Horizon 2020 research and innovation programme under grant agreement No 769872. 
%\begin{figure}[H]
 %\centering
 %\includegraphics[width=0.3\columnwidth]{eu.png}
 %\caption{EU logo.}
 %\Description{EU logo}
 %\label{fig:eulogo}
%\end{figure}

%
% The next two lines define the bibliography style to be used, and the bibliography file.
\bibliographystyle{ACM-Reference-Format}
\bibliography{petra-ws}

\end{document}